\RequirePackage{lineno}
\documentclass[prb,twocolumn,aps,amsmath,amssymb,floatfix]{revtex4}
\usepackage{graphicx} 
\usepackage{bm} 
\usepackage{ulem,epstopdf}
\usepackage{color,graphicx,pstricks}
\begin{document}

\title {Interfacial-antiferromagnetic-coupling driven magneto-transport
  properties in ferromagnetic superlattices}

\author{Sandip Halder$^1$, Subrat K. Das$^2$, and
  Kalpataru Pradhan$^{1,}$\footnote{kalpataru.pradhan@saha.ac.in}}

\affiliation{$^1$Theory Division, Saha Institute of Nuclear Physics,
  A CI of Homi Bhabha National Institute, Kolkata 700064, India\\
  $^2$SKCG Autonomous College, Paralakhemundi, Odisha 761200, India}

\begin{abstract}
We explore the role of interfacial antiferromagnetic interaction in coupled
soft and hard ferromagnetic layers to ascribe the complex variety of
magneto-transport phenomena observed in $La_{0.7}Sr_{0.3}MnO_3/SrRuO_3$
(LSMO/SRO) superlattices (SLs) within a one-band double exchange model using
Monte-Carlo simulations. Our calculations incorporate the magneto-crystalline
anisotropy interactions and super-exchange interactions of the constituent
materials, and two types of antiferromagnetic interactions between Mn and Ru
ions at the interface: (i) carrier-driven and (ii) Mn-O-Ru bond super-exchange
in the model Hamiltonian to investigate the properties along the hysteresis
loop. We find that the antiferromagnetic coupling at the interface induces
the LSMO and SRO layers to align in anti-parallel orientation at low temperatures.
Our results reproduce the positive exchange bias of the minor loop
and inverted hysteresis loop
of LSMO/SRO SL at low temperatures as reported in experiments.
In addition, conductivity calculations show that the carrier-driven
antiferromagnetic coupling between the two ferromagnetic layers steers the
SL towards a metallic (insulating) state when LSMO and SRO are aligned in
anti-parallel (parallel) configuration, in good agreement with the experimental
data. This demonstrate the necessity of carrier-driven antiferromagnetic
interactions at the interface to understand the one-to-one correlation 
between the magnetic and transport properties observed in experiments.
For high temperature, just below the ferromagnetic $T_C$ of SRO, we unveiled
the unconventional three-step flipping process along the magnetic
hysteresis loop.
We emphasize the key role of interfacial antiferromagnetic coupling
between LSMO and SRO to understand these multiple-step flipping
processes along the hysteresis loop. 
\end{abstract}

\maketitle

\section {Introduction}
Transition metal oxides, particularly those with the perovskite
structure, are materials of great interest due to both, their basic
physics~\cite{goodenough,khomskii} and potentiality
in technological applications~\cite{zubko,hwang,mannhart}. They
exhibit a wide variety of collective, coupled complex magnetic behaviours
in bulk form~\cite{khomskii,fiebig,wang}. Novel physics emerges from
bilayers of two such materials which are
believed to be among the potential optimal heterostructures for future
technological spintronics applications~\cite{feng,wu1,yu1,ramirez}. It
is generally agreed upon that it is the interface that decides the
coupling between the layers and the overall properties of the
heterostructure~\cite{bruno,garcia,hoffman,he1,murthy}. Magnetic order at
interfaces often drives the magnetism of the constituent materials in
bilayers~\cite{bhattacharya,hwang}. Some of the key features
that originates due to the magnetic reconstructions at the interface are
unusual tunnelling magnetoresistance~\cite{anil}, in-verse spin-Hall
effect~\cite{emori,wahler},
exchange-bias~\cite{giber,zhou,ke1,meiklejohn1} etc.

The exchange bias (EB) effect is one of the most widely studied
interface phenomena observed in many magnetic materials and
heterostructures~\cite{nogues,magnin,nogues1,guo,rana,nogues2,gruyters}. 
Stronger interfacial interaction can change the magnetic response
of a heterostructure dramatically as compared to its constituent
counterparts~\cite{bruno1,koon,murthy}. In case of a coupled soft/hard
ferromagnetic heterostructures with very different coercive fields 
the magnetization of the softer ferromagnet (FM) can selectively be `twist' wrt the 
harder FM during the hysteresis loop~\cite{binek}. The interfacial
interaction has the ability to shift the magnetic hysteresis loop,
making it asymmetric about zero applied field. This
feature, known as the EB effect is extensively used to pin the
magnetization of the hard FM. The shift of the magnetization
loop in the direction of (opposite to) applied bias field is
referred as positive (negative) exchange bias. In case of positive
exchange bias the interfacial exchange interaction is believed to
be antiferromagnetic~\cite{ke1}. EB induced pinning of magnetization
of one magnetic layer has significant potential applications in spin
valves~\cite{radu,binek}, magnetic recording read
heads~\cite{zhang1}, magnetic random access memory circuits~\cite{huijben},
giant magnetoresistive sensors~\cite{bibes2} etc.

Thin film heterostructures of hard and soft FMs are of great interest for
EB realization, which also lead to the
appearance of the very interesting phenomenon of inverted hysteresis loop
(IHL)~\cite{ziese,saghayezhian,ghising}. In the hysteresis loop, generally,
the remanence is found to be positive with magnetization $M$ oriented along
the applied field $h$ when one reduces the field strength to zero from
it's saturation value. On the other hand, in case of IHL the
magnetization is found to be aligned along the opposite direction to
the applied field $h$, when still $h>0$, while decreasing the field. As a result
an IHL show cases a negative coercivity and negative remanence. Such an
anomalous behavior of magnetization $M(h)$ curve is observed in amorphous
Gd-Co films~\cite{esho}, bulk ferrimagnet of composition
$Er_2CoMnO_6$~\cite{banerjee}, exchange coupled multilayers~\cite{bloemen},
hard/soft multilayers~\cite{sabet,fullerton}, and single domain particles
with competing anisotropies~\cite{das1}.

Ferromagnetic half-metallic manganites are considered to as good
candidates for engineering spintronic devices. Particularly, heterostructures
comprised of LSMO as one of the constituent materials such as
$LSMO/SRO$~\cite{das2,behera}, $LSMO/BiFeO_3$~\cite{singh}
and $LSMO/La_{0.7}Sr_{0.3}CoO_3$~\cite{li} are extensively studied
in last decades. Especially, LSMO and SRO are attractive materials due
to their epitaxial growth and lattice-matched heterostructures which show
several interface driven interesting magnetic
phenomena~\cite{padhan1,padhan2,thota,zhang2,ziese1}. LSMO is a well studied
half-metallic ferromagnet~\cite{park,bowen,urushibara}, where as SRO is a
rare 4$d$ based oxide having ferromagnetic ordering~\cite{kanbayasi,grutter}.
In recent past the temperature dependence of magnetization reversal mechanism
has also been investigated in perovskite ferromagnetic oxide's superlattice,
LSMO/SRO~\cite{ziese2}. The interplay of interlayer exchange coupling and
magneto-crystalline anisotropy results in an inverted hysteresis loop at
low temperatures. In addition, at higher temperature (close to the $T_C$
value of SRO) the superlattice showed an unconventional triple flip
mechanism (LSMO$\uparrow$ SRO$\uparrow$ to LSMO$\uparrow$ SRO$\downarrow$
to LSMO$\downarrow$ SRO$\uparrow$ to LSMO$\downarrow$ SRO$\downarrow$),
where the SRO layer switches first on reducing the magnetic field
from saturation value~\cite{ziese2}. This ferrimagnetic SL configuration
flips its alignment for magnetic field of opposite polarity and later both
the layers align along the external magnetic field. So, underlying flipping
mechanism in LSMO/SRO SL makes it a suitable model system for
theoretical investigations.

In this work our aim is to present a qualitative understanding of the
underlay mechanism of experimentally observed EB, IHL and the unconventional
triple-flip behaviour of LSMO/SRO SLs~\cite{ziese2} emphasizing the
role of interlayer antiferromagnetic couplings~\cite{ziese1,lee}. In
order to investigate these interesting temperature dependent magnetic along with
transport properties we construct a model Hamiltonian for the LSMO/SRO like
SL systems and employ the Monte-Carlo technique based on travelling cluster
approximation (for handling large size systems)~\cite{kumar1,pradhan1}.
We observe the EB and IHL at low temperature and the unconventional triple
flip behaviour of magnetic hysteresis loop at high temperature similar to
the experiments. We find that a stronger inter-layer antiferromagnetic
coupling is necessary to realize the multiple flips nature of the
hysteresis curve. The antiferromagnetic interactions at the interface
gains strength both from carrier-driven and bond-driven interactions
between Mn and Ru ions. But,the carrier-driven antiferromagnetic interaction
at the interface is necessary to understand the one-to-one correspondence
between magnetic and transport properties observed in LSMO-SRO SLs.

The organization of the paper is as follows: In next section we present
the electronic structures and the relevant properties of the constituent
materials (LSMO and SRO), and the nature of the interfacial coupling in
their SL configuration. In section III  we introduce a suitable model
Hamiltonian and the methodology to solve the SL systems, while section IV
establishes the parameter values for the constituent bulk materials by
reproducing their essential properties qualitatively. In section V we start
our SL calculations with exhibiting the fact that the LSMO and SRO layers 
align antiferromagnetically at low temperatures that triggers an
insulator-metal transition with decreasing the temperature. In section VI
and VII we present the underlay mechanism of the exchange bias and the
inverted hysteresis loop, respectively, as observed in low-temperature
experiments. Then we discuss the more unconventional high-temperature
three-step switching process of the magnetic hysteresis loop in LSMO/SRO
superlattices and emphasize the role of interfacial antiferromagnetic
coupling in section VIII. Section IX summarizes our key findings.

\begin{figure}[!t]
\centerline{
\includegraphics[width=8.5cm,height=6.30cm,clip=true]{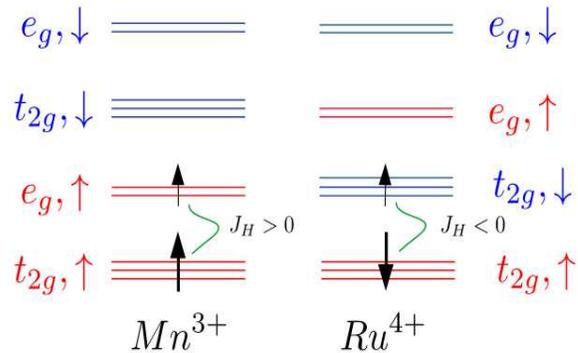}}
\caption{
Schematically shown the energy band diagrams of
$3d$ orbitals of $Mn^{3+}$ ions in LSMO (in left panel)
and $4d$ orbitals of $Ru^{4+}$ ions in SRO (in right panel).
The Hund's coupling constant $J_H$ between the majority
core spins and the minority spin is positive for $Mn^{3+}$ ions and negative
for $Ru^{4+}$ ions. The spacing of energy levels are not drawn to scale.
}
\label{sch_band}
\end{figure}

\begin{figure}[!t]
\centerline{
\includegraphics[width=8.5cm,height=10.3cm,clip=true]{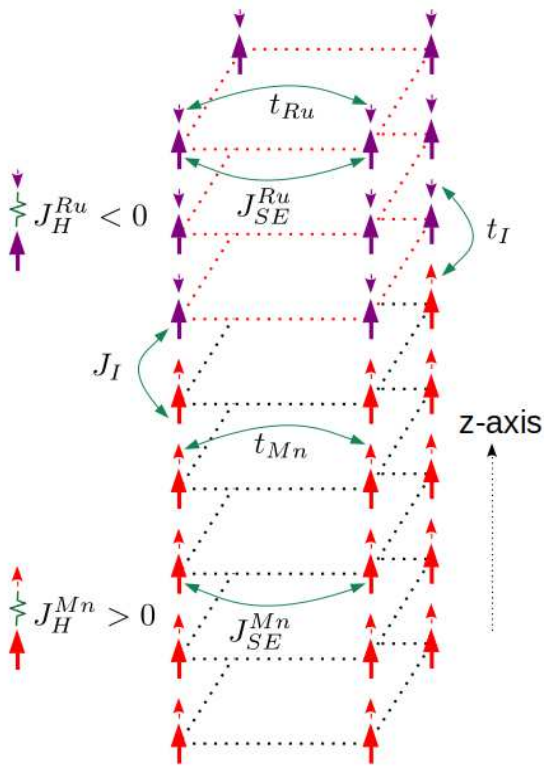}}
\caption{
Schematic of the lattice model for LSMO/SRO superlattice based on Mn and Ru ions
showing the super-exchange interactions ($J_{SE}^{Mn}$, $J_{SE}^{Ru}$, $J_I$),
hopping parameters ($t_{Mn}$, $t_{Ru}$, $t_I$) and
Hund's couplings ($J_H^{Mn}$, $J_H^{Ru}$).}
\label{sch_model}
\end{figure}

\section{Electronic Structures of Constituent Materials}

Here, we discuss the relevant properties of the constituent layers in
LSMO/SRO superlattice. LSMO is the optimally doped compound of $LaMnO_3$
with Sr in place of La, where three 3$d$ valence electrons of $Mn$ ion are
occupied in $t_{2g}$ orbitals and the $e_{g}$ orbitals remain partially
filled~\cite{dagotto,tokura}. The average electron density $n$=0.7 is
defined on the basis of occupancy of $e_{g}$ electrons. Three electrons in
the $t_{2g}$ orbitals form a large core spin. We consider it as classical
spin~\cite{dagotto1} ${\bf |S|}$ and fix ${\bf |S|=1}$, which is coupled
to the itinerant $e_{g}$ electrons spin via the Hund's
coupling~\cite{dagotto1,zener,yunoki,pradhan}. Delocalization of $e_{g}$
electrons results in ferromagnetic metallic state in the double
exchange limit. Here, we assign Hund's coupling to be large and positive,
where itinerant electrons are aligned along the core spins ($3\uparrow, 1\uparrow$)
as shown in Fig.~\ref{sch_band}. This high spin configuration forms a
total moment $\sim 3.7\mu_{B}$ as reported in experiments~\cite{coey}.
Experimental observations show that LSMO has relatively a high ferromagnetic 
transition temperature ($T_C$ $\sim$ 360 K), negligible magnetic anisotropy
and low coercive field~\cite{steenbeck}.

Coming to the next material, $SrRuO_{3}$ (SRO) is believed to be an
itinerant ferromagnet with finite density of states at the Fermi
level~\cite{singh1}. Here, Ru is in the $Ru^{4+}$ state and all 4$d^4$
electrons are occupied in $t_{2g}$ orbitals ($3\uparrow, 1\downarrow$)
forming a low spin configuration state of total moment $2\mu_{B}$ which
is close to the experimentally observed value $\sim 2.2\mu_{B}$~\cite{ning}.
Other experimental reports suggest that the $Ru^{4+}$ ion moment in SRO
varies wildly depending upon the growth condition and found to be between
1.0-1.6 $\mu_{B}$~\cite{koster,bushmeleva,allen,klein,rondinelli}.
It is also known that compressively strained films exhibit enhanced
saturated magnetic moments~\cite{grutter}. The lower value of magnetic moment
indicates the itinerant nature of the $t_{2g}$ electrons. But, the
coexistence of localized and itinerant magnetism can not be ruled out
completely~\cite{dodge,shai,kim}. In our model Hamiltonian approach we
consider $3\uparrow$ electrons as localized classical spin ${\bf |S|}$
which is Hund's coupled to the $\downarrow$ itinerant electrons with an
average system electron density $n=0.5$ which give rise to an effective
moment $\sim$$2.5\mu_{B}$~\cite{ning,chang}. The sign of Hund's coupling
constant of low spin state $t_{2g}$ electrons in SRO is of opposite sign
to that we assigned in LSMO (schematically shown in Fig.~\ref{sch_band})).
Although both LSMO and SRO have metallic conductivity at low temperatures
SRO is relatively a bad metal~\cite{toyota,klein1} and the ferromagnetic
transition temperature of SRO $T_C^{SRO}\sim$160 K~\cite{longo,koster}
which is much smaller than that of LSMO. Experimental
observations show that SRO has larger magnetic anisotropy and higher
coercive field in comparison to LSMO~\cite{kanbayasi1,koster,ziese3}.
As a result, LSMO is a most preferred candidate for soft magnetic materials
when coupled with hard ferromagnet like SRO, which has a relatively larger
coercive field~\cite{ziese,ziese1}.

Based on above facts, where itinerant electrons are polarized opposite
(parallel) to the direction of localized $t_{2g}$ core spins in SRO (LSMO),
it is clear that the itinerant electrons prefer antiferromagnetic alignment
(over the ferromagnetic alignment) of Mn and Ru $t_{2g}$ core spins at the
interface to facilitate the hopping and gain kinetic energy~\cite{chang}.
As a result the interfacial hopping drive the system to be less resistive in
the antiferromagnetic configuration between the constituents ferromagnetic
layers as compared to the ferromagnetic or paramagnetic configuration, in
agreement with experiments~\cite{takahashi}. This shows that carrier-driven
antiferromagnetic interaction is necessary to understand the one-to-one
correspondence between magnetic and transport properties of the LSMO/SRO
SLs along the hysteresis loop. It is worthwhile to note that in addition
to this carrier-driven coupling, interfacial Mn-O-Ru bonds also generate
a stronger antiferromagnetic super-exchange coupling at the
interface~\cite{ziese1,lee} which is also needs to be incorporated.

\section{Model Hamiltonian and Methods}

With the backdrop of the electronic structure and electron density
we construct a reference one band classical Kondo lattice model
Hamiltonian~\cite{yunoki,pradhan,chakraborty} 
in three dimensions to investigate the LSMO/SRO heterostructures as follows:

\[H= -t\sum_{<ij>,\sigma}^{}(c_{i\sigma}^{\dagger}c_{j\sigma}
+c_{j\sigma}c_{i\sigma}^{\dagger})\,-\, J_{H}\sum_{i}{\bf S}_{i}.
{\bf \sigma}_{i}\]
\[+ J\sum_{<ij>}{\bf S}_{i}.{\bf S}_{j} - A_{aniso}\sum_{i}(S_{i}^z)^2\]
\[-\mu \sum_{i\sigma} c_{i\sigma}^{\dagger}c_{i\sigma}
+\sum_{i\sigma}\epsilon_{i} c_{i\sigma}^{\dagger}c_{i\sigma.}\]
\noindent
Here, the operator $c_{i\sigma}$ ($c_{i\sigma}^{\dagger}$)
annihilates (creates) an itinerant electron with spin
$\sigma_{i}$ at site $i$. $J_{H}$ is the Hund's coupling between the
$t_{2g}$ spins $\bf S_{i}$ and the itinerant electron spin ${\bf \sigma}_{i}$.
We treat $\bf S_{i}$ to be classical variable and fix $\bf |S_{i}|=1$.
$J$ is the antiferromagnetic super-exchange between the classical
{\bf $S_{i}$} spins and $A_{aniso}$ is the strength of magneto-crystalline
anisotropy. Chemical potential $\mu$ is tuned to set the average electron
density of the overall system. $\epsilon_{i}$ is the onsite potential,
essential for layered systems to keep the electron densities of both layers
at their desired values. This term can be neglected for bulk systems.
We perform our calculations in an external magnetic field $h$ by adding
the Zeeman coupling $-{\bf h}\cdot\sum_i {\bf S}_i$ to the Hamiltonian.

We apply the exact diagonalization scheme to the itinerant electrons
in the configuration of the fixed background of classical spins.
The classical variables are annealed by Monte-Carlo procedure
at each site where the proposed update is accepted or rejected by using
metropolis algorithm starting with the random initial configuration.
At each temperature we use 2000 system sweeps for annealing and in each
sweep visit every lattice site sequentially and update the system.
We measure physical parameters like magnetization after thermalizing
the system at each 10 sweeps to avoid illicit self-correlation in the
data. In order to access larger system size we use travelling cluster
approximation~\cite{kumar1,pradhan1} based Monte-Carlo scheme. We set
up our model Hamiltonian calculations for m-LSMO/n-SRO heterostructure
in 3D (N=$10\times10\times8$). Here, $m$($n$) represents the number of
LSMO (SRO) planes with $m+n=8$. The size of the TCA cluster is taken to
be $4\times 4\times 8$. Now onward, we assign different SL structures
as $m/n$ SL, where $m$ LSMO planes constitute the LSMO layer
and $n$ SRO planes constitute the SRO layer. In order to establish
the bulk properties of individual LSMO and SRO layers we use
$8\times8\times8$ system size.

\section{Two sets of parameter values to mimic properties of bulk LSMO and SRO}

In order to capture the essential physics of individual LSMO and SRO layers
qualitatively, first we have to explore and find out two different sets of
parameter values comprising of $t$, $J_{H}$, $J_{SE}$ and $A_{aniso}$. 
Keeping the basic properties of the constituent materials in mind, 
first we consider to build up the parameter space for the LSMO. We set
$t_{Mn}=1$ and calculate all the observables in the unit of $t_{Mn}$.
In order to mimic the ferromagnetic metallic state the Hund's coupling
constant is set in the double-exchange limit, $J_{H}^{Mn}=24$. We add a
small antiferromagnetic super-exchange interaction putting $J_{SE}=0.01$
and neglect the magneto-crystalline anisotropy ($A_{aniso}^{Mn}=0$)
which is much smaller than the SRO system.

Now, for SRO the parameters must be chosen in such a way that its physics
relative to LSMO remain intact at least qualitatively e.g. the transition
temperature $T_C^{SRO} < T_C^{LSMO}$, the coercive field $H_C^{SRO} > H_C^{LSMO}$
etc. Hence, we chose the hopping parameter $t_{Ru}=0.5$, the Hund's coupling
$J_{H}^{Ru}=-12$ and the magneto-crystalline anisotropy interaction
$A_{aniso}^{SRO}=0.05$. The modality by which we assign the negative sign to
$J_{H}^{Ru}$ is already discussed earlier (please see the discussion related to
Fig.~\ref{sch_band}). A finite $A_{aniso}^{SRO}$ is essential to capture the 
higher value of coercive field $H_C$ in SRO compared to LSMO at low temperatures.
We neglect any kind of super-exchange interaction in SRO. The electron
densities are already fixed from the electronic structures of both materials,
$n\sim 0.7$ for LSMO and $n\sim 0.5$ for SRO as outlined earlier.
We tabulate the parameter sets for LSMO-like and SRO-like materials in
Table.1.

\begin{table}[!t]
\centering
\begin{tabular}{|l|l|l|l|l|}
\hline
{\bf System}& {\bf Parameters to mimic LSMO and SRO like}\\    
{}& {\bf systems}\\ \hline
{LSMO}& $n=0.7$, $J_{H}^{Mn}=24$, $t_{Mn}=1$, $J_{SE}^{Mn}=0.01$, $A_{aniso}^{LSMO}=0$ \\ \hline
{SRO}& $n=0.5$, $J_{H}^{Ru}=-12$, $t_{Ru}=0.5$, $J_{SE}^{Ru}=0$, $A_{aniso}^{SRO}=0.05$ \\ \hline
\end{tabular}
\caption{Various parameters to model LSMO and SRO like systems. For LSMO/SRO SLs we set
$J_I=0$ and $t_I=0.75$ unless otherwise mentioned.}
\label{para_set}
\end{table}

\begin{figure}[!t]
\centerline{
\includegraphics[width=8.5cm,height=6.30cm,clip=true]{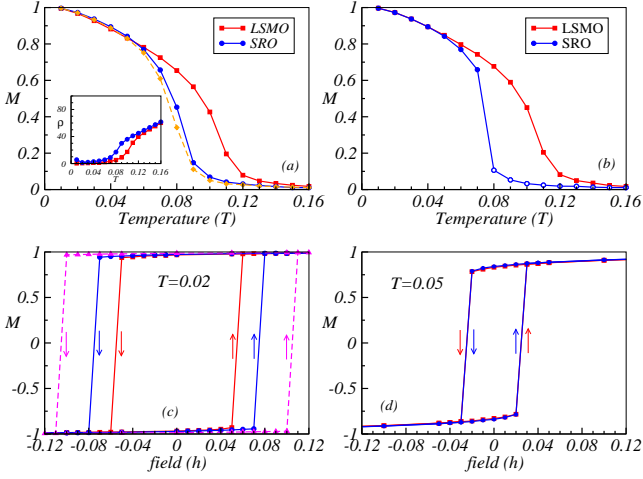}}
\caption{
Displays the bulk properties of individual LSMO and SRO systems computed
using the parameter sets listed in Table-1.
(a) and (b) present the magnetization M vs temperature T results
with $A_{aniso}^{SRO}$=0.05 in the whole temperature range and
with $A_{aniso}^{SRO}$=0.05 (0.00) below (above) $T_C^{SRO}$, respectively.
In both cases $T_{C}^{LSMO} > T_{C}^{SRO}$ with a reduced
value of $T_{C}^{SRO}$ for $A_{aniso}^{SRO}$=0.0 above $T_C^{SRO}$ case.
Also shown the M vs T of SRO-like materials for $J_H=-9$ using the
dashed line. Most of our calculations are performed for $T < T_C^{SRO}$
and we employ the parameters used in (a) to model LSMO and SRO in rest
of our calculations. Inset shows that both LSMO and SRO are
metallic at low temperatures.
(c) and (d) show the M vs applied field $h$ hysteresis loops
of LSMO-like and SRO-like materials at low temperature $T=0.02$
and moderate temperature $T=0.05$, respectively.
The arrows indicate the field sweep direction.
Clearly, Coercive field of SRO $H_C^{SRO} > H_C^{LSMO}$ at $T=0.02$,
but comparable to each other at $T=0.05$. Legends are same in panels
(a), (c) and (d). Also, we show the M-h hysteresis loop for
SRO like materials for $A_{aniso}^{SRO}$=0.1 in (c) using dashed
lines. 
}
\label{bulk}
\end{figure}

Next, the task is to calculate and establish the essential properties of
LSMO-like and SRO-like materials separately using the parameter sets given
in Table-1. We present the magnetization $M$ $=1/N\sum_{i}S_{z}^{i}$ data
in Fig.~\ref{bulk}(a), where $N$ is number of sites and $S_{z}$ is the $z$
component of the classical spins. The magnetization with temperature graphs
show that the ground state is ferromagnetic for both the materials with
$T_C^{LSMO} > T_C^{SRO}$. In experiments, the $T_C^{LSMO}$ is measured to
be as large as twice of $T_C^{SRO}$. We agree that our calculations do not
reflect this fact quantitatively. It is important to note here that the
magneto-crystalline anisotropy is mainly a low temperature effect. So taking
$A_{aniso}^{SRO}=0$ above $T_C$ and $A_{aniso}^{SRO}=0.05$ below $T_C$
(keeping all other parameters fixed as in (a)) we present the calculations for
SRO in Fig.~\ref{bulk}(b). Now, the ferromagnetic $T_C^{SRO}$ is reduced and
consequently the gap between the $T_C$ of the two systems widens, which is a
better match to the experimental results. In this work, most of the calculations
are carried out below the ferromagnetic $T_C$ of SRO ($T_C^{SRO}$) and
high temperature means $T \lesssim T_C^{SRO}$, similar to experiments. In that
spirit we use the parameter space employed in (a) for both LSMO and SRO to avoid
confusions. In the inset of Fig.~\ref{bulk}(a) we show the resistivity for both
the systems. We obtain the resistivity by calculating the $dc$ limit of the
conductivity as determined by the Kubo-Greenwood formula~\cite{mahan-book,kumar2,pradhan2}.
Our calculations show that both systems are metallic at low temperature which
agrees with the experimental results.

Next, we compare the coercive field $H_C$ of both the systems in Fig.~\ref{bulk}(c). 
Our calculations shows that $H_C^{LSMO} < H_C^{SRO}$ at low temperature ($T=0.02$)
giving rise to a combination of hard and soft ferromagnets as required for
modelling the superlattice (SL). On the other hand, $H_C^{LSMO}$ is comparable to
$H_C^{SRO}$ at intermediate temperatures [see ~\ref{bulk}(d)] as we move close to
$T_C^{SRO}$. So, the two sets of parameters that we assigned to LSMO-like and
SRO-like materials capture qualitatively the essential physics that is required
to investigate their superlattices. Hence, we call them LSMO and SRO in our further
analysis. In addition, in Fig.~\ref{bulk}(a), using the dotted line,
we have shown that the M-T data is no different if one uses $J_H^{Ru}=-9$. We
also confirm that $H_C^{SRO}$ remains same for both $J_H^{Ru}$ = -9 and -12 (not
shown in figure). What would happen to the M-h curve if we take a larger anisotropy
constant $A_{aniso}^{SRO}=0.1$ instead of $A_{aniso}^{SRO}=0.05$, particularly at low
temperatures? Obviously the $H_C^{SRO}$ gets enhanced further as shown using
the dotted line in Fig.~\ref{bulk}(c). 

In experiments, during the magnetic hysteresis measurements (at low temperature)
of LSMO/SRO SLs magnetization of the SRO layer is pinned up to moderate field
strength for positive field cooled system, whereas the
LSMO layer switches its orientation during field sweep. This is a consequence
of higher coercive field of SRO as compared to LSMO. In addition, a positive exchange
bias is also observed in experiments below the ferromagnetic $T_C^{SRO}$ which is believed
to be a consequence of anti-ferromagnetic inter-layer coupling~\cite{ziese1}.
In fact, this antiferromagnetic interlayer coupling plays a significant role in
transport and magnetic properties of the LSMO/SRO SLs. Here, it is an
interesting scenario where interlayer coupling turns out to be antiferromagnetic 
although the constituent layers themselves have dominant ferromagnetic
interactions among the core spins. This is explained using the transformed
interfacial bond arrangements and resulting interfacial charge transfer.
Density functional calculations suggest that the bond angle Mn-O-Ru at the
interface drives an antiferromagnetic super-exchange coupling~\cite{ziese1,lee},
whose strength would be different from the super-exchange interaction within
either layer. The impact of interfacial electronic charge transfer is two fold: 
(i) it modifies the antiferromagnetic super-exchange interaction and
(ii) induces an antiferromagnetic interaction via carrier driven process
at the interface, as discussed earlier.
This charge carrier mediated coupling will very much depend upon the modified
hopping parameter $t_I$ at the interface. In fact, the strength of charge
transfer driven antiferromagnetic coupling would be larger for $t_I=t_{Mn}=1$
as compared to that of $t_I=t_{Ru}=0.5$.

\begin{figure}[!t]
\centerline{
\includegraphics[width=8.5cm,height=6.30cm,clip=true]{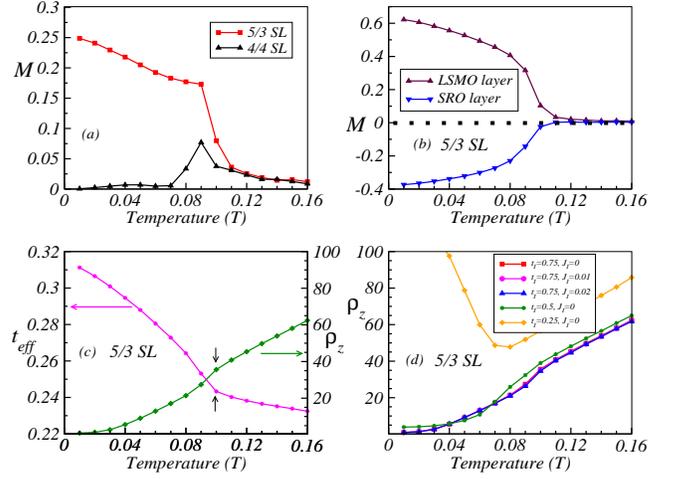}}
\caption{
Magnetic and transport properties of LSMO/SRO SLs:
(a) Variation of M with T for the 5/3 SL and 4/4 SL showing the sign of
onset of antiferromagnetic interfacial coupling at $T \sim 0.1$.
(b) M vs T of individual LSMO and SRO layers for 5/3 SL
confirms the fact that at low temperatures both the layers
are aligned ferromagnetically with antiferromagnetically coupled
at the interface.
(c) Variation of out-of-plane resistivity $\rho_z$ and
effective hopping parameter $t_{eff}$ with temperature for the
5/3 SL showing an insulator-metal transition at $T=0.1$.
The sharp change in $t_{eff}$ at $T=0.1$
is due to the gain in the kinetic energy of the carriers in the SL.
This corroborates the fact that antiferromagnetic interfacial coupling
(shown in (b)) drives the system towards a metallic phase.
(d) Variation of $\rho_z$ vs $T$ for different combinations of
interfacial hopping parameter and interfacial super-exchange
interaction ($t_I$, $J_I$) values depict that suppressed carrier
hopping across the interface ($t_I$=0.25) negates the insulator-metal
transition.
}
\label{tc_sl}
\end{figure}

\section{Establishing the antiparallel alignment of LSMO and SRO layers}

It is generally assumed that the inter-layer coupling decides the overall
properties of the heterostructures. So, we start the SL structure
calculations with understanding the nature of interfacial interaction 
in 5/3 and 4/4 superlattices (SLs). These notions of SL structures are
already discussed. To maintain the desired electron densities in both the
layers one has to be chosen the relative on-site potential $\epsilon$ accordingly. 
$\epsilon$ takes the value $\bigtriangleup$ (0) in LSMO (SRO) layer.  
And, we found that for $\bigtriangleup=10.7$ the LSMO and SRO layers maintain
the electron densities 0.7 and 0.5, respectively, in both 5/3 SL and 4/4 SL
systems. LSMO and SRO layers are coupled at the interface via hopping
parameter $t_{I}$ and the super-exchange interaction $J_{I}$. We set
$t_{I}$=0.75 and $J_{I}=0$ unless otherwise mentioned. This will
help us to first analyze the effects of the carrier driven interfacial
antiferromagnetic coupling. In subsequent calculations the consequence of
super-exchange coupling is emphasized wherever necessary.

We measure the magnetization of the 5/3 and the 4/4 SLs by cooling from
high temperature $T=0.16$ in a very small external field $h=0.002$
[see Fig.~\ref{tc_sl}(a)]. In both cases the magnetization starts to increase
at $T \sim 0.11$, which is attributed to the ferromagnetic alignment of the high
$T_C$ constituent LSMO layer. At $T \sim 0.09$ the magnetization decreases for
4/4 SL, whereas the slope of magnetization curve changes abruptly in case of
5/3 SL. These results indicate that the SRO layer starts to align
ferromagnetically but in opposite orientation to LSMO layer at $T=0.09$. 
In order to verify this fact we plot the magnetization of embedded LSMO and
SRO layers separately for 5/3 SL in Fig.~\ref{tc_sl}(b), which we found to
be consistent with the results presented in Fig.~\ref{tc_sl}(a). So,
the antiferromagneic interaction at the interface drives the layers to
align antiferromagnetically at low temperatures. Now the question arises:
does this antiferromagnetic alignment at $T$ $\sim 0.09$ leads to a metallic
state according to our hypothesis, presented earlier, where we argued that
antiferromagnetic alignment at the interface facilitate the delocalization
of charge carriers. To check this we calculated the out-of-plane resistivity
$\rho_z (T)$ and presented in Fig.~\ref{tc_sl}(c). The resistivity depicts
an insulator-metal transition around $T\sim 0.1$ (indicated by black arrow)
where both layers start to align antiferromagnetically with each other as shown
in Fig.~\ref{tc_sl}(a) and (b).  This ensures that the LSMO/SRO SL systems with
antiferromagnetic interfacial coupling are metallic in nature. In order to establish the
correspondence between resistivity and gain in kinetic energy we calculate
the effective hopping~\cite{white,mondaini}
\[t_{eff} = \langle\sum_{j,\sigma}(c_{j,\sigma}^{\dagger}c_{j+z,\sigma}
+c_{j+z,\sigma}c_{j,\sigma}^{\dagger})\rangle \]
of 5/3 SL, where angular bracket represents the expectation value.
The $t_{eff}$ vs temperature shows a sharp change at
$T=0.1$ indicating that gain in kinetic energy of the SL system gets enhanced
at the same temperature where both LSMO and SRO layers start to align
antiferromagnetically. So our $t_{eff}(T)$ and $\rho_{z}(T)$ results
compliment each other.

In order to establish the fact that the carrier-driven antiferromagnetic
coupling between the two ferromagnetic layers steers the SL towards a
metallic (insulating) state when LSMO and SRO are aligned in anti-parallel
(parallel) configuration, as seen in experiments, we calculate the magnetic
and transport properties using different combinations of $t_I$ (interfacial
hopping parameter) and $J_I$ (interfacial super-exchange interactions).
For varying $J_I$ with a fixed $t_I=0.75$ we did not find any variation
in the resistivity curve as shown in Fig.~\ref{tc_sl}(d). But for small
values of $t_I$ (with fixed $J_I=0$) the system does not go to a metallic
state at low temperatures although LSMO and SRO layers align
antiferromagnetically (not shown in figure). This shows that a reasonable
interfacial hopping that also drives the antiferromagnetic interaction is
necessary to understand the magneto-transport properties of LSMO/SRO SLs.

\begin{figure}[!t]
\centerline{
\includegraphics[width=8.5cm,height=6.30cm,clip=true]{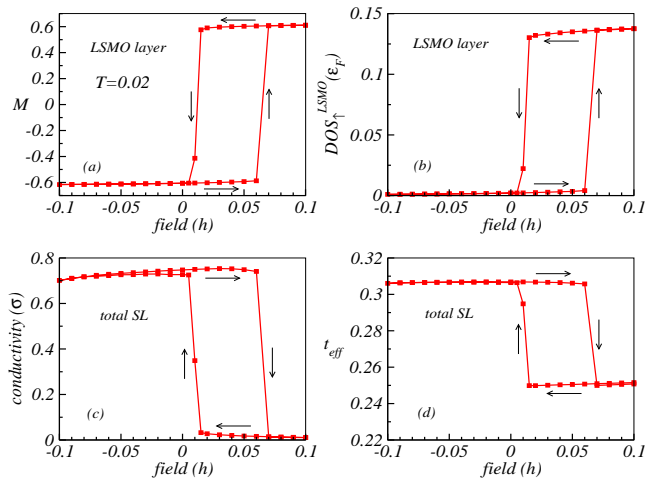}}
\caption{
Focuses on the minor hysteresis loops (the hysteresis of only LSMO layer)
of 5/3 SL at low temperature $T=0.02$. Here, the applied field is
restricted within a region such that only the magnetization of
LSMO flips its direction, leaving the SRO layer pinned along the up
direction. The arrows indicate the field sweep direction.
(a) Shows the positive exchange bias of the soft magnetic layer LSMO with
a shift of the M-h hysteresis loop to the right of the zero field axis.
(b) Displays the hysteresis of spin-up density of states of LSMO layer
at the Fermi level ($D_{\uparrow}^{LSMO}(\varepsilon_{F})$) which follows
magnetic hysteresis curve shown in (a).
(c) Hysteresis of conductivity shows that the SL is more conducting
for oppositely oriented LSMO and SRO layers as compared to their
parallel configuration.
(d) Hysteresis of effective hopping parameter $t_{eff}$ follows
conductivity loop which corroborate the fact that the carriers are
more mobile in antiparallel configuration as compared to the parallel
configuration of both layers across the interface.
}
\label{eb_T02}
\end{figure}

\section{Positive Exchange Bias}

We have established the essential properties of LSMO, SRO and their SLs 
qualitatively. Now, we come to the key calculation of the article, where
we measure the magnetic hysteresis M-h loop. We start with 5/3 SL at
low temperature $T=0.02$. First, we cool down the SL from high temperature
to $T=0.02$ under an external magnetic field $h=0.1$ and then measure
the M-h loop. The field cycle of $h$ loop is restricted to $\pm$0.1, 
so that the magnetization of the SRO layer remains in the up direction
and does not flip at all. We will see in next section that SRO flips its
direction beyond $h=-0.1$. Here, the aim is to analyze the minor hysteresis
loop of the SL, which is the full hysteresis loop of the embedded
LSMO layer. We have presented the variation of magnetization of LSMO layer
in Fig.~\ref{eb_T02}(a). For $h=+0.1$ LSMO and the SRO layers are aligned
ferromagnetically along the field direction. As we decrease the magnetic
field strength  the magnetization of the LSMO layer flips its direction at
a small +ve field (LSMO$\uparrow$ SRO$\uparrow$ to LSMO$\downarrow$ SRO$\uparrow$).
Primarily, there are two reasons of this flip: 
(i) Coercive field of LSMO is smaller than SRO ($H_C^{LSMO} < H_C^{SRO}$)
as shown in Fig.~\ref{bulk}(c) 
and (ii) the interfacial antiferromagnetic coupling who wins over the
low field strength in aligning the LSMO layer in its favour. While sweeping
back the field from $h = -0.1$, a higher field strength ($h=0.07$)
is required to overcome the antiferromagnetic inter-layer coupling and flip
back the LSMO magnetization along the field direction. As a result, the
hysteresis loop of the LSMO layer shifts towards the positive field axis
giving rise to a positive exchange bias. In addition, we have calculated
the spin-resolve density of states (DOS) at the Fermi level
($\varepsilon_{F}$) of the LSMO layer and plot the spin up density of states 
$D_{\uparrow}^{LSMO}(\varepsilon_{F})$ in Fig.~\ref{eb_T02}(b), which
perfectly follows the LSMO magnetic hysteresis loop shown in
Fig.~\ref{eb_T02}(a). This shows that if the carriers are perfectly
aligned to the core spins the DOS of the soft ferromagnet can be used as
a parameter to track the magnetization flipping in LSMO/SRO like SLs.

We already discussed that the carriers will be more delocalized
(localized) across the interface in SL structure when the magnetic
moments in LSMO and SRO layers are aligned antiferromagnetically
(ferromagnetically) with each other. To check this fact, we have
calculated the $dc$ conductivity throughout the magnetic hysteresis
loop and found that the SL is more conducting in the anti-parallel
configuration as compared to the parallel configuration of LSMO and
SRO layers as shown in Fig.~\ref{eb_T02}(c). We also show that the
hysteresis loop of the effective hopping $t_{eff}$
[see Fig.~\ref{eb_T02}(d)] is very similar to conductivity hysteresis
loop, corroborating the fact that carriers are more mobile across the
interface in anti-parallel configuration (LSMO$\downarrow$ SRO$\uparrow$)
as compared to parallel orientation (LSMO$\uparrow$ SRO$\uparrow$).

\begin{figure}[!t]
\centerline{
\includegraphics[width=8.5cm,height=6.30cm,clip=true]{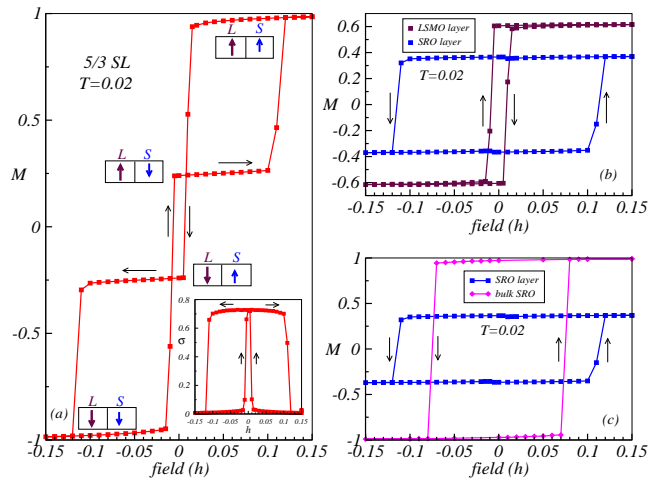}}
\caption{
Magneto-transport properties along the magnetic hysteresis measurements
for 5/3 SL at $T=0.02$:
(a) The complete magnetic hysteresis curve shows a
centrally inverted hysteresis loop with negative coercivity
and negative remanence as observed in the experiment~\cite{ziese2}.
Relative magnetization of LSMO and SRO layers (LSMO as L and SRO as S) is
illustrated schematically.
Inset shows that the conductivity in anti-parallel configuration
is larger than that of the parallel configuration along the hysteresis loop.
(b) Separate plots of magnetic hysteresis curves of the embedded
SRO and LSMO layers depict that the SRO magnetization behaves
conventionally, in contrast to the LSMO magnetization which is inverted.
(c) The magnetic hysteresis of the embedded SRO layer is compared with
that of the free standing bulk system [replotted as in Fig.~\ref{bulk}(c)].
The coercive field $H_C$
of embedded SRO layer is larger than that of the bulk SRO due to the
interfacial coupling, which resist the layer to flip at $H_C^{SRO(bulk)}$.
}
\label{ihl_T02}
\end{figure}

\section{Inverted Hysteresis loop}

We now investigate the full hysteresis loop of the 5/3 SL at low temperature
$T=0.02$. Here and in later sections all the calculations are carried out
after cooling down the SL in the external field $h_{cooling}$= + 0.15,
which is higher than the $H_C^{SRO}$. At this forward saturation field
strength the magnetizations of both LSMO and SRO layers are aligned parallel
to the external field. It is also important to remember that
the SRO layer (LSMO layer) acts as a hard (soft) ferromagnet at $T=0.02$.
So, by decreasing the field from $h$=+0.15 towards $h$=-0.15, first the LSMO
layer reverses its magnetization at small but finite positive fields giving
rise to a negative remanence. Then the SRO layer flips its magnetization
at a larger external magnetic field of opposite polarity (at $h\sim -0.11$)
as shown in Fig.~\ref{ihl_T02}(a). Now, traversing the field back from
$h$=-0.15 the LSMO magnetization flips first at a small negative field
resulting in a negative remanence and then SRO flips at a higher positive field
(at $h\sim 0.11$) strength, giving rise to a central inverted hysteresis
loop (IHL) which is very similar to the experimental results~\cite{ziese2}.
For clarity, we have also plotted the magnetic hysteresis of individual
LSMO and SRO layers embedded in SL in Fig.~\ref{ihl_T02}(b). The SRO layer
shows a conventional hysteresis loop, but LSMO layer depicts an inverted loop.
This is because the interfacial antiferromagnetic coupling steers LSMO
layer to align opposite to field direction by overcoming the external field
energy to establish the antiferromagnetic configuration giving rise to a
negative remanence. So, the full M-h hysteresis loop at low temperature
reveals the striking magnetization reversal process that is absent in the
minor hysteresis loop of the SL that we presented in the previous section.
The SL system is found to be metallic (insulating) along the hysteresis
loop where LSMO and SRO are aligned in anti-parallel (parallel) configuration
as shown in the inset of Fig.~\ref{ihl_T02}(a). This fact is in good
agreement with experimental results~\cite{takahashi}.

\begin{figure}[!t]
\centerline{
\includegraphics[width=8.5cm,height=6.30cm,clip=true]{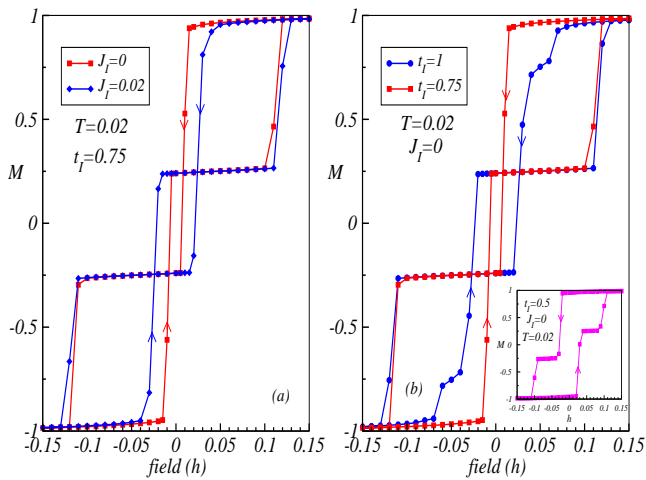}}
\caption{
Comparison of magnetic hysteresis loops of 5/3 SLs at $T=0.02$
with (a) variation of the interfacial super-exchange interaction $J_I$
at fixed value of interfacial hopping parameter $t_I=0.75$ and
(b) variation of $t_I$ at fixed $J_I=0$. Both $J_I$ and $t_I$ enhance
the interfacial coupling strength. So, the width of the central IHL
broadened with increasing $J_I$ and/or $t_I$.
Inset: IHL is absent for the parameters set $(t_I, J_I)=(0.5,0)$,
which indicates that a reasonable strength of the interfacial
antiferromagnetic coupling is essential to realize the inverted
hysteresis feature.
}
\label{ihl_j}
\end{figure}

In addition, we compared the magnetic hysteresis of the embedded SRO layer
with the bulk SRO system at $T=0.02$ in Fig.~\ref{ihl_T02}(c). The coercive
field of SRO in SL is larger than that of the bulk SRO,
$H_{C}^{SRO(SL)} > H_{C}^{SRO(bulk)}$. The earlier flipping of LSMO layer
as shown in Fig.~\ref{ihl_T02}(a) favours the interfacial antiferromagnetic
coupling which opposes the SRO layer to flip in the field direction up to
a certain field strength that is larger than $H_{C}^{SRO(bulk)}$. As a result,
relatively a large field of opposite polarity is required to flip
the SRO layer as compared to its bulk counterpart. 

\begin{figure}[!t]
\centerline{
\includegraphics[width=8.5cm,height=6.30cm,clip=true]{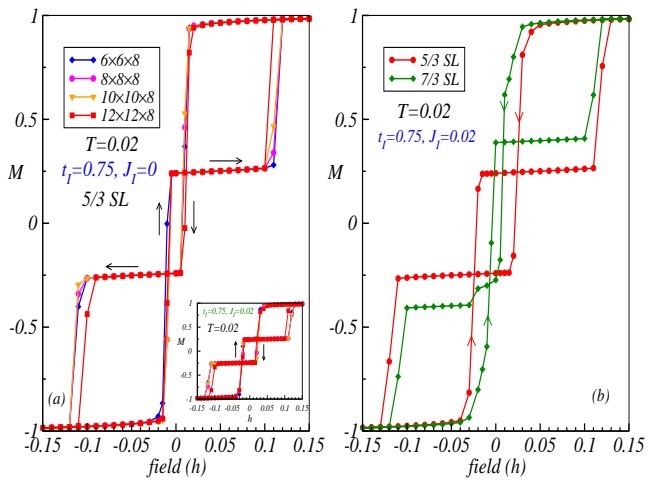}}
\caption{
The stability of the IHL with different system sizes:
(a) Magnetic hysteresis along with the IHLs for four different
system sizes are seen to be almost indistinguishable from each other.
All systems are belong to the 5/3 SL structure.
Inset: The IHLs are also very stable for $J_I=0.02$.
(b) Shows the shrinking of the width of the IHL for 7/3 SL structure
(system size: $10\times10\times10$) as compared to the 5/3 SL due to
thicker LSMO layer (7 planes), which flips its magnetization at a lower
applied field strength as compared to the thinner LSMO layer (5 planes).
}
\label{ihl_L}
\end{figure}

It is earlier mentioned that primarily there are two sources of inter-layer
antiferromagnetic interaction: (i) carrier-mediated antiferromagnetic
coupling and (ii) Mn-O-Ru bond driven super-exchange coupling ($J_I$).
In all calculations presented till now, we incorporated only the first
type of interaction among the core spins at the interface for analyzing
the hysteresis loop. Now, we consider both interactions and investigate
the inverted hysteresis loop in Fig.~\ref{ihl_j} to emphasize the effect
of Mn-Ru direct super-exchange coupling. We compare the M-h hysteresis loops  
for different super-exchange coupling strengths $J_I$=0 and 0.02
at fixed $t_I=0.75$ at
low temperature $T=0.02$ in Fig.~\ref{ihl_j}(a). The central inverted part
of the hysteresis loop is seen to be more prominent for $J_I$=0.02.
The reason is the enhancement of overall (effective) antiferromagnetic
interaction at the interface that facilitate the rotation of magnetization
(from up to down direction) of the LSMO layer at a larger positive applied
field. The carrier-mediated antiferromagnetic coupling can also be enhanced
(for fixed $J_I$=0) by increasing the inter-layer hopping parameter $t_I$ to
1 instead of 0.75, which can be understood from the inverted part of the
hysteresis loop shown in Fig.~\ref{ihl_j}(b). The inter-layer coupling
strength decreases considerably in case of suppressed carrier hopping
($t_I =0.5$), which results in a conventional hysteresis loop without any
inverted part (see inset of Fig.~\ref{ihl_j}(b)). 
These results clearly show the crucial role of the 
interfacial antiferromagnetic coupling 
in generating the central inverted hysteresis loop.

Next, we demonstrate the stability of the inverse hysteresis loop for
different system sizes. In Fig.~\ref{ihl_L}(a) shows the M-h hysteresis
loops for four different system sizes, namely N=$6\times6\times8$, $8\times8\times8$,
$10\times10\times8$ and $12\times12\times8$ at low temperature $T=0.02$.
Here, we have modified only the size of the planes keeping the thickness
of the SLs and individual layers fixed. So all these calculations are for
different sizes but each one of them falls under the 5/3 SLs group.
Interestingly, the overall hysteresis loops including the central 
inverted parts are hardly distinguishable from each other for all four cases.
This shows that the magnetic properties, we discussed for the
system size N=$10\times10\times8$, 
remains very robust without any significant size effect. We have also shown that
the magnetic hysteresis loop remains unaffected by varying the size of the system
for $J_I = 0.02$ (see the inset of Fig.~\ref{ihl_L}(a)).

Switching of LSMO moments within small magnetic field strengths results in 
the IHL for 5/3 SL at $T=0.02$ and $J_I$=0, discussed in Fig.~\ref{ihl_T02}.
Tuning on the inter-layer super-exchange antiferromagnetic interaction
($J_I$=0.02) the central IHL becomes more prominent (see Fig.~\ref{ihl_j}).
What will happen to the central IHL if one varies the thickness of the LSMO
layer, keeping the SRO thickness fixed?  To check this, we considered:
(i) 7/3 SL with a $10\times10\times10$ lattice and 
(ii) 9/3 SL with a $10\times10\times12$ lattice for $J_I$=0.02. 
Keeping the thickness of SRO layer fixed if we increase 
the LSMO layer thickness it is clearly visible that the area of 
the central inverted loop part shrinks (see Fig.~\ref{ihl_L}(b)). 
And, if we further increase it then the IHL vanishes and 
conventional loop reappears for 9/3 SL (not shown in the figure).
In this case the magnetization of the LSMO layer flips below $h=0$ 
resulting in the disappearance of IHL.
This is because the thick LSMO layer does not prefer to get rotated easily
(rotation is based on the interfacial antiferromagnetic coupling) as compared
to the thin LSMO layer against the magnetic field energy. These results
indicate the importance of relative thicknesses of the constituent layers to
realize the IHL in SLs.

\begin{figure}[!t]
\centerline{
\includegraphics[width=8.5cm,height=6.30cm,clip=true]{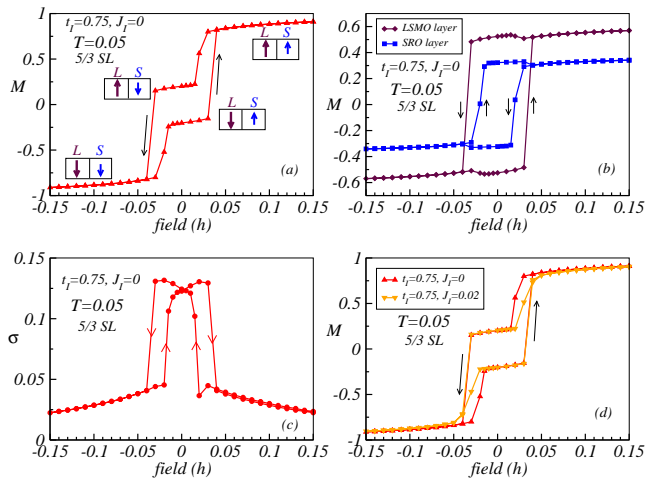}}
\caption{
Presents the magneto-transport properties for 5/3 SL at intermediate
temperature $T=0.05$, where the coercive fields of bulk LSMO and SRO are
comparable to each other (see Fig.~\ref{bulk}(d)). We set $J_I=0$.
(a) M-h hysteresis loop shows a two-step switching process, but without
any central inverted part.
(b) Clarifies the switching process with separately plotted the
M-h data of embedded LSMO layer. (b) along with (a) ensure that the
magnetization of SRO layer flips first which is followed by LSMO layer.
In fact, embedded SRO layer shows inverted hysteresis feature.
(c) Shows that conductivity in parallel
orientation remains smaller than that of the anti-parallel configuration
along the M-h hysteresis loop.
(d) Additional interfacial super-exchange coupling $J_I=0.02$
does not change the qualitative features of the two-step flipping
process of M-h hysteresis loop.
}
\label{2s_T05}
\end{figure}

\section{Unconventional three-step flipping process along the hysteresis loop}

The temperature dependent hard or soft nature of LSMO and SRO results in  
unconventional and interesting magnetic hysteresis loops in LSMO/SRO SLs at
low temperatures. We have calculated and discussed, in detail, the hysteresis
loop at low temperature where the coercive fields of LSMO and SRO layers are
quite different from each other. What if the coercive fields of two materials
are comparable to each other? How does the magnetic hysteresis loop will behave
in such a scenario? For this we have to study the M-h loops at high temperatures.
Before going to analyze high temperature calculations it would be
interesting to explore the intermediate temperature regime where the coercive
fields of the constituent layers are already comparable to each other
(see Fig.~\ref{bulk}(d) for $T=0.05$). Hence, we plot the M-h magnetic hysteresis
curves for 5/3 SL at $T=0.05$ in Fig.~\ref{2s_T05}(a), which shows a two-step
process as in the low temperature case. In high field strength both layers are
aligned along the field direction. Sweeping from the forward to the reverse
saturation field the SRO layer flips first  at a small +ve field creating a
ferrimagnetic SL magnetization and ultimately both layers align along the field
direction (LSMO$\uparrow$ SRO$\uparrow$ to LSMO$\uparrow$ SRO$\downarrow$
to LSMO$\downarrow$ SRO$\downarrow$) at saturation field. This flipping scenario
is more clear in Fig.~\ref{2s_T05}(b) where
we have plotted the magnetization of the individual LSMO and SRO layers. It is
clear that LSMO layer is the second one to switch its magnetization direction.
In fact, the flips of only SRO layer depict an inverted hysteresis loop similar
to the LSMO layer presented in Fig.~\ref{ihl_T02}(b) for low temperatures.  
The inverted part is not visible for whole SL structure at $T=0.05$ due to the
smaller effective magnetic moment in SRO layer which happens to flip first
in this case. So, to visualize the central inverted hysteresis loop it is necessary 
that the higher moment layer should flip first and give a negative remanence. 
The $dc$ conductivity in Fig.~\ref{2s_T05}(c) shows that the conductivity in
parallel configuration remains smaller compared to the anti-parallel
configuration, similar to low temperature case. The two-step
switching process as a function of the magnetic field remains intact even
modifying the inter-layer interaction by tuning on the super-exchange
mediated coupling to $J_I=0.02$ (see Fig.~\ref{2s_T05}(d)).

\begin{figure}[!t]
\centerline{
\includegraphics[width=8.5cm,height=6.30cm,clip=true]{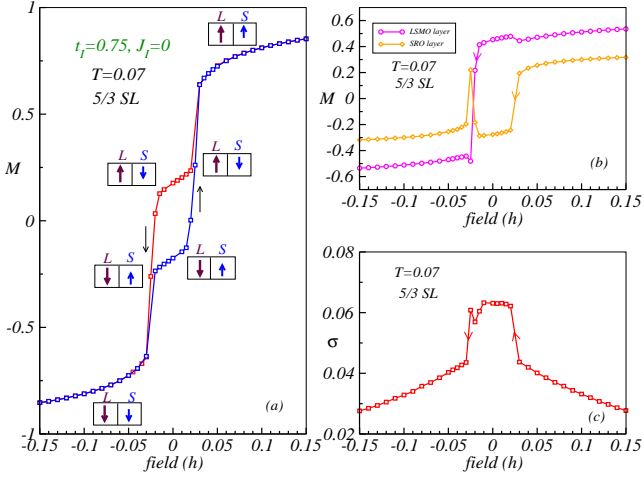}}
\caption{
Displays the magneto-transport properties of 5/3 SL at high temperature
$T=0.07$ (close to $T_C^{SRO}$). (a) M-h hysteresis measurements depict
the unconventional three-step magnetization flipping process,
qualitatively similar to the experimental results~\cite{ziese2}.  
(b) The magnetization switching of the embedded individual LSMO and SRO
layers
during M-h cycle confirm the three-step flipping.
(c) Conductivity measurements along the hysteresis loop
show that layers in the anti-parallel configuration are more conducting
than that in the parallel orientation. But, their difference is reduced
at high temperature as compared to low temperature.
}
\label{3s_T07}
\end{figure}

At both low and intermediate temperature a two-step switching process is
observed. LSMO (SRO) flips first at $T=0.02$ ($=0.05$) and generates a
ferrimagnetic configuration of the magnetic moments in the SL in the
intermediate-step of M-h hysteresis curve. But at high temperature, close
to $T_C^{SRO}$, an unconventional three-step switching process
(LSMO$\uparrow$ SRO$\uparrow$ to LSMO$\uparrow$ SRO$\downarrow$
to LSMO$\downarrow$ SRO$\uparrow$ to LSMO$\downarrow$ SRO$\downarrow$)
is observed
in the experiment~\cite{ziese2}. Here, the SRO layer flips first at a positive
low field followed by a switching of the overall ferrimagnetic SL magnetization
in negative low fields and a further switching of the SRO layer to align along
the field direction. In order to understand this complicated unconventional
three-step process we study the $M-h$ hysteresis curve at $T=0.07$ in
Fig.~\ref{3s_T07}(a). This temperature is just below the $T_C$ of SRO, similar
to the experimental set up. During the field sweep starting from $h$=+0.15,
sweeping from the forward to the reverse saturation field, the SRO layer
switches its direction first, as in case of $T=0.05$, making the ferrimagnetic
SL. This ferrimagnetic configuration changes its overall direction at
$h \sim -0.05$. Then, further going beyond this field strength both the
layers orient in the field direction. The three-step flipping process is
also observed during traversing sweep which emphasizes the stability of the
second ferrimagnetic configuration (LSMO $\downarrow$ SRO$\uparrow$) along
the hysteresis loop.
This three-step flipping process is depicted in Fig.~\ref{3s_T07}(b)
where we have plotted the magnetization of the individual layers (plotted
only for sweeping from the forward to the reverse saturation field). The
carrier mediated antiferromagnetic interaction at the interface for fixed
$J_I=0$ plays the vital role in stabilizing the both ferrimagnetic phases
which we will discuss soon. Due to the
high operating temperature value the saturation magnetization of both the
layers are much smaller than the saturation magnetization values observed
for earlier calculations. Also the $dc$ conductivity in parallel
configuration is found to be only marginally smaller than the anti-parallel
configuration (see Fig.~\ref{3s_T07}(c)).

\begin{figure}[!t]
\centerline{
\includegraphics[width=8.5cm,height=6.30cm,clip=true]{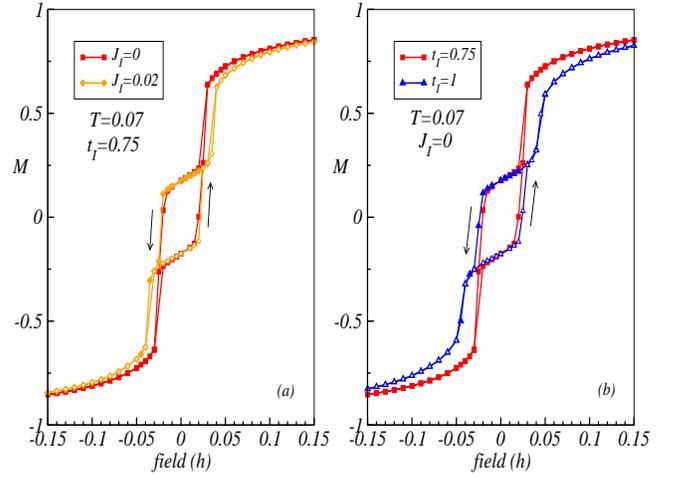}}
\caption{
Demonstrates the effect of interfacial antiferromagnetic coupling
on the three-step flipping process of 5/3 SL at high temperature $T=0.07$.
In (a) the interfacial super-exchange interaction $J_I$ is varied
with keeping the interfacial hopping parameter fixed at $t_I=0.75$,
and in (b) $t_I$ is varied with fixed $J_I=0$. The interfacial
coupling strength increases with increasing $J_I$ and/or $t_I$.
So in both cases the three-step flipping process is more
prominent for higher values of $J_I$ and $t_I$.
}
\label{3s_j}
\end{figure}

In the three-step switching process (shown in Fig.~\ref{3s_T07}(a)), although
the reversal of ferrimagnetic SL configuration is observed, the second
ferrimagnetic configuration (LSMO$\downarrow$ SRO$\uparrow$) withstands over
a very narrow magnetic field window. In order to show that the interfacial
antiferromagnetic interaction plays a significant role in stabilizing this
ferrimagnetic phases we incorporated the super-exchange interaction and
calculated the magnetic hysteresis loop of the SL at $T=0.07$. The
three-step switching process is found to be more prominent, particularly
the second ferrimagnetic configuration (LSMO$\downarrow$ SRO$\uparrow$)
for $J_{I}$=0.02 as
shown in Fig.~\ref{3s_j}(a). This indicates
that the LSMO layer prefers to orient along the field direction at a particular
small negative field value for which ferrimagnetic SL configuration reverses
to maintain the antiferromagnetic interaction between the two layers at
the interface. This is also supported from the results obtained from
varying the inter-layer hopping parameter $t_I$ with fixed $J_I=0$. As the
carrier driven antiferromagnetic coupling at the interface increases with
$t_I$ the three-step magnetic flipping process is more prominent for
$t_I$=1 as compared to $t_I$=0.75 (see Fig.~\ref{3s_j}(b)). But surprisingly
the magnetic switching remains a two-step process for $t_I$=0.5 due to
insufficient strength of the interfacial antiferromagnetic coupling.
Here, if we incorporate a large super-exchange interaction ($J_I=0.05$) then only
the three-step flipping process is recovered (not shown in figure). These
results establish that a reasonable strength of interfacial antiferromagnetic
coupling (comprised of carrier driven and bond driven) between LSMO and SRO
is required to realize the three-step switching process at high temperature
as seen in the experiments.

\section{CONCLUSIONS}

In this work we investigated the interfacial antiferromagnetic interaction
driven complex magneto-transport phenomena observed in LSMO/SRO
superlattices (SLs) within a one-band double exchange model using a
Monte-Carlo technique based on travelling cluster approximation.
We have considered the appropriate magneto-crystalline-anisotropy interaction
and super-exchange interactions terms of LSMO and SRO (the constituent
materials) in the model Hamiltonian. Also, our calculations incorporate:
(i) the carrier-driven and (ii) the bond-mediated super-exchange 
interfacial antiferromagnetic interactions between
Mn and Ru to demonstrate the vital role of the interfacial coupling in deciding
the magneto-transport properties along the hysteresis loop. Our conductivity
calculations show that the anti-aligned core spins at the interface prompt
the carrier hopping between both layers to gain kinetic energy and
consequently steers the SL system towards a metallic (less resistive) phase.
On the other hand, the SL is found to be in a less metallic or insulating
state when the LSMO and SRO layers are aligned in the same direction.
Hence, the system undergoes metal-insulator or insulator-metal transition
upon varying the temperature and/or the applied magnetic field.
Interestingly, by invoking the antiferromagnetic coupling at the interface,
we explained the exchange bias effect and inverted hysteresis loop at
low temperatures, and unconventional three-step flipping mechanism at
high temperatures (close to ferromagnetic $T_C$ of SRO),
which are in good agreement with the experimental results.
In addition, our calculations establish that the carrier-driven
antiferromagnetic interaction is one of the necessary ingredient to
understand the one-to-one correlation between the magnetic and the
transport properties of LSMO/SRO SLs observed in experiments.

\section*{ACKNOWLEDGMENT}
We acknowledge use of the Meghnad2019 computer cluster at SINP.

\end{document}